\documentclass[12pt]{book}

\usepackage[dvips]{graphicx,color}
\usepackage{makeidx,tsukuba}

\makeauthorindex
\makeindex

\begin{document}

\BookTitle{\itshape The 28th International Cosmic Ray Conference}
\CopyRight{\copyright 2003 by Universal Academy Press, Inc.}
\pagenumbering{arabic}

\chapter{
Dissipation of Hydromagnetic Waves on Energetic Particles: Impact on
Interstellar Turbulence and Cosmic Ray Transport}

\author{%
%
%
V.S. Ptuskin,$^{1,2}$ F.C. Jones,$^3$ I.V. Moskalenko,$^{3,4}$ and V.N. 
Zirakashvili$^1$\\
{\it (1) IZMIRAN, Troitsk, Moscow region 142190, Russia\\
(2) University of Maryland,College Park, MD 20742, USA\\
(3) NASA/Goddard Space Flight Center, Greenbelt, MD 20771, USA\\
(4) University of Maryland, Baltimore, MD 21250, USA}
}

\section*{Abstract}

The diffusion of galactic cosmic rays (CR) is considered. It is anticipated
that the nonlinear MHD cascade sets the power-law spectrum of turbulence 
from
the principle scale $100$ pc to much smaller scales. We found that the
dissipation of waves due to the resonant interaction with energetic 
particles
may terminate the cascade at less than $10^{13}$ cm. The  shape of CR
diffusion coefficient that was found may explain the observed peaks of 
secondary to primary
nuclei ratios at a few GeV/n.

\section{Introduction}

The galactic CR have high energy density and they can not always be treated 
as
test particles moving in given magnetic fields. In particular, the 
stochastic
acceleration of CR by MHD waves is accompanied by the damping of waves. The
wave damping causes the change of the wave spectrum that in turn affects the
particle transport. Thus in principle the study of CR diffusion may need a
selfconsistent consideration. We shall see that the CR action on 
interstellar
turbulence should be taken into account at energies below $10$ GeV/n. The
implementation of this effect in a full scale numerical simulations of CR
propagation in the Galaxy was fulfilled by Moskalenko et al. (this 
Conference).

\section{Equations for Cosmic Rays and Interstellar Turbulence}

The steady state transport equation that describes the CR diffusion in the
interstellar medium is of the form, e.g. [1]:
\begin{equation}
-\nabla D\nabla\Psi-\frac{\partial}{\partial 
p}p^{2}K\frac{\partial}{\partial
p}p^{-2}\Psi=q,
\end{equation}
where $\Psi(p,\mathbf{r},t)$ is the particle distribution function on 
momentum
normalized on CR number density as $N_{cr}(p)=\int_{p}^{\infty}dp\Psi$,
$D(p,\mathbf{r})$ is the spatial diffusion coefficient, $K(p,\mathbf{r})$ is
the diffusion coefficient on momentum, and $q(p,\mathbf{r})$ is the source
term. If needed, the supplementary terms which describe the particle energy
losses, nuclear fragmentation, and radioactive decays \ can be added to eq.
(1). It is useful to introduce the diffusion mean free path $l=3D/v$.

It is assumed that the particle diffusion is due to the resonant 
wave-particle
interaction. The diffusion mean free path is determined by the equation [1]:
$l=r_{g}B^{2}\left(  4\pi k_{res}W(k_{res})\right)  ^{-1}$, where $v$ is the
particle velocity, $r_{g}=pc/(ZeB)$ is the particle Larmor radius in the
average magnetic field $B$, $k_{res}=1/r_{g}$ is the resonant wave number, 
and
$W(k)$ is the spectral energy density of waves defined as $\int 
dkW(k)=\delta
B^{2}/4\pi$ ($\delta B$ is the random magnetic field, $\delta B\ll B$). The 
CR
diffusion coefficient has the scaling $D\propto v(p/Z)^{a}$ for the power 
law
spectrum $W(k)\propto1/k^{2-a}$. The diffusion on momentum is roughly
described by the formula $K=p^{2}V_{a}^{2}\left(  a(4-a)(4-a^{2})D\right)
^{-1}$, where $V_{a}$ is the Alfven velocity.

In spite of the great progress in magnetic hydrodynamics, we do not yet have
the well developed theoretical description of interstellar turbulence that
would allow one to calculate $W(k)$ in different astrophysical conditions, 
see
review [3]. It is quite possible that there are two almost independent
nonlinear cascades of waves in the magnetized plasma where the thermal
pressure approximately equals the magnetic field pressure, e.g. [2]. The
cascade of Alfven waves (and the slow magnetosonic waves) leads to the
Kolmogorov spectrum $W(k)\propto k^{-5/3}$, and the cascade of fast
magnetosonic waves leads to the Kraichnan spectrum $W(k)\propto k^{-3/2}$ in
the inertial range of wave numbers where the dissipation is absent. Below we
consider the interaction of CR with these two cascades separately.

\textit{A. Kolmogorov type cascade. }In its simplified form, the steady 
state
equations for waves with a nonlinear transfer in $k$-space can be written as
\begin{equation}
\frac{\partial}{\partial k}\left(  C_{A}k^{2}\sqrt{kW(k)\left(  4\pi
\rho\right)  ^{-1}}W(k)\right)  =-2\Gamma_{cr}(k)W(k)+S_{A}\delta(k-k_{L}).
\end{equation}
$k\geq k_{L}$, $\rho$ is the gas density (see [7,8], the original theory of
Kolmogorov (1941) was developed for the incompressible liquid without 
magnetic
field). The l.h.s.\ of eq. (3) describes the nonlinear cascade from small 
$k$
to large $k$, $C_{A}$ is a constant, and approximately equal to 0.3 
according to
the simulations [10]. The r.h.s.\ of eq. (3) includes the wave damping on CR
and the source term, which works on the main scale $1/k_{L}=100$ pc
and describes the generation of turbulence by supernova bursts, stellar 
winds,
and superbubbles expansion. In the limit of negligible damping $\Gamma_{cr}%
=0$, the solution of eq. (2) gives the Kolmogorov scaling $W(k)\propto
k^{-5/3}$.

The equation for wave amplitude attenuation on CR is [1]:
\begin{equation}
\Gamma_{cr}(k)=\pi e^{2}V_{a}^{2}\left(  2kc^{2}\right)  ^{-1}\int
_{p_{res(k)}}^{\infty}dp\,p^{-1}\Psi(p),
\end{equation}
where $p_{res}(k)=ZeB/ck$. The solution of eqs (2), (3) allows finding the
wave spectrum and the determination of the mean free path, which is
\begin{equation}
l(p)=l_{Ko}(p)\left[  1-2\pi^{3/2}V_{a}p^{1/3}l_{Ko}^{1/2}(p)\left(
3C_{A}B^{2}r_{g}^{1/2}\right)  ^{-1}\int_{p}^{p_{L}}dp_{2}p_{2}^{2/3}%
\int_{p_{2}}^{\infty}\frac{dp_{1}}{p_{1}}\Psi(p_{1})\right]  ^{-2}.
\end{equation}
Here $l_{Ko}$ is the diffusion mean free path calculated for a Kolmogorov
spectrum without regard of wave damping, and $p_{L}=p_{res}(k_{L})$. The
second term in brackets describes the modification of the mean free path due
to the damping of short waves.

\textit{B. Iroshnikov-Kraichnan cascade. }The simplified equation for waves
reads simular to eq. (3) but with the different l.h.s.: $\frac{\partial
}{\partial k}\left(  C_{M}k^{3}\left(  \rho V_{a}\right)  ^{-1}W^{2}%
(k)\right)  $, where approximately $C_{M}=1$. At $\Gamma_{cr}=0$, this gives 
the
spectrum $W(k)\propto k^{-3/2}$ first found in [4,6]. Using the same 
procedure as in
the case A, one can obtain:
\begin{equation}
l(p)=l_{Kr}(p)\left[  1-\pi V_{a}p^{1/2}l_{Kr}(p)\left(  2C_{M}B^{2}%
r_{g}\right)  ^{-1}\int_{p}^{p_{L}}dp_{2}p_{2}^{1/2}\int_{p_{2}}^{\infty}%
\frac{dp_{1}}{p_{1}}\Psi(p_{1})\right]  ^{-1}.
\end{equation}
Here $l_{Kr}$ is the diffusion mean free path calculated for a power law
Kraichnan spectrum without regard of wave damping on CR.

As the most abundant species, the CR protons mainly determines the wave
dissipation. Their distribution function $\Psi(p)$ should be used to 
calculate
$l(p)$ by the simultaneous solution of eqs (1) and (4) in the case A, and
eqs (1) and (5) in the case B. The diffusion mean free path for other nuclei
is $l(p/Z)$.

Let us estimate the effect of wave damping at $1$ GeV where approximately 
$l=1$
pc. The CR energy density is $1$ eV/cm$^3$, $V_{a}=10$ km/s, $B=3$
$\mu$G. The second term in brackets in eq.(4) equals $5\times10^{-2}$,
whereas the second term in brackets in eq.(5) equals $10$. We conclude that
the Kolmogorov type cascade is not much affected by the damping on CR. The
Iroshnikov-Kraichnan cascade is significantly affected, and this should lead
to the modification of CR transport at energies less than about $10$ GeV.

\section{Simple Selfconsistent Solution}

To demonstrate the effect of wave damping, we consider a simple case of
one-dimensional diffusion with the source distribution 
$q=q_{0}(p)\delta(z)$,
$q_{0}\propto p^{-\gamma_{S}}$ (that corresponds to the infinitely thin
disk of CR sources located at $z=0$) and the flat CR halo of height $H$, see
[5]. Let us assume that stochastic reacceleration does not essentially 
change
the CR spectrum during the time of CR exit from the Galaxy, i.e. one can set
$K=0$ in eq. (1). The solution of eq. (1) in the galactic disk is then
$\Psi(p)=3q_{0}(p)H\left(  2vl(p)\right)^{-1}$. We consider the
Iroshnikov-Kraichnan cascade and simplify eq. (5) using the approximation
$\int_{p_{2}}^{\infty}dp_{1}p_{1}^{-1}\Psi(p_{1})=\Psi
(p_{2})/(\gamma_{s}+0.5)$. This allows one to find the self consistent 
diffusion
mean free path for protons:
\begin{equation}
l(p)= l_{Kr}(p)\exp\left(  3\sqrt{\pi}eH\left(  8(\gamma
_{s}+0.5)C_{M}\sqrt{\rho}c\right)  ^{-1}\int_{p}^{p_{L}}dp_{1}q_{0}%
(p_{1})v^{-1}(p_{1})\right).
\end{equation}
The form of eq. (6) is close to that needed to explain the peaks in ratios 
of
secondary to primary nuclei in CR at a few GeV/n. Note, that the escape 
length
(the  grammage) that determines the production of secondaries is equal to
$X=3\mu H/(2l)$, where $\mu$ is the surface gas density of the galactic disk 
[5].

\section{Conclusion}

The damping on CR terminates slow Iroshnikov-Kraichnan cascade
but probably has no impact on the Kolmogorov-type cascade in the
interstellar medium. The estimates were made for the turbulence level that
provides the empirical value of CR diffusion coefficient. This
finding offers a new explanation of the peaks in the secondary/primary 
ratios
at a few GeV/n: the amplitude of short waves is small because of the
damping and thus the low energy particles rapidly exit the Galaxy and almost 
do not produce secondaries.
Another explanation, e.g. [9], that the peaks
are produced by CR reacceleration ($K\neq0$) on an undisturbed 
Kolmogorov-type
spectrum remains as a viable alternative. The quantitative analysis of this 
problem
is presented by Moskalenko et al. at this Conference.

Many aspects of this topic including the structure of turbulence at the main 
scale
(the theory predicts that $S_{A}\gg S_{M}$), the anisotropy in
$k$-space (the waves propagates at large angles to local magnetic field in 
the
Alfvenic turbulence and this should increase the diffusion coefficient
compared to (2)), and the consequences for
the interpretation of data on interstellar turbulence remain
beyond the scope of the present short paper.

\textbf{Acknowledgments.} The work was supported by NASA Astrophysical 
Theory
Program grants. It was also supported by RFBR grant at IZMIRAN.

\section{References}

%
%
%
%
%
%
%

\re
1.\ Berezinskii V.S. et al. 1990, Astrophysics of Cosmic Rays (North 
Holland,
Amsterdam)\re
2.\ Cho J., Lazarian A. 2002, astro-ph/0205282\re
3.\ Goldstein M.L., Roberts D.A., Matthaeus W.H. 1995, ARA\&A 33, 283\re
4.\ Iroshnikov P. 1963, Sov. Astron. 7, 566\re
5.\ Jones F.C. et al.\ 2001, ApJ 547, 264\re
6.\ Kraichnan R.\ 1965, Phys. Fluids 8, 1385\re
7.\ Landau L.D., Lifschitz E.M.\ 1987, Fluid Mechanics (Pergamon, Oxford)\re
8.\ Norman C.A., Ferrara A.\ 1996, ApJ 467, 280\re
9.\ Seo E.S., Ptuskin V.S.\ 1994, ApJ 431, 705\re
10.\ Verma M.K. et al.\ 1996, JGR 101, 21619

\endofpaper
\end{document}